\documentclass[]{aastex631}

\received{December 10, 2024}
\revised{May 5, 2025}
\accepted{\today}
\submitjournal{Astrophysical Journal}

\shorttitle{Data-driven radiative hydrodynamics simulations}
\shortauthors{Keller}

\graphicspath{{./}}

\begin{document}

\title{
  Data-driven radiative hydrodynamics simulations of the solar photosphere using physics-informed neural networks: proof of concept
 }

\email{ckeller@nso.edu}
\author[0000-0002-0786-7307]{Christoph U. Keller}
\affiliation{National Solar Observatory \\
3665 Discovery Dr \\
Boulder, CO 80303, USA}

\begin{abstract}
Current, realistic numerical simulations of the solar atmosphere reproduce observations in a statistical sense; they do not replicate observations such as a movie of solar granulation. Inversions on the other hand reproduce observations by design, but the resulting models are often not physically self-consistent. Physics-informed neural networks (PINNs) offer a new approach to solving the time-dependent radiative hydrodynamics equations and matching observations as boundary conditions. PINNs approximate the solution of the integro-differential equations with a deep neural network. The parameters of this network are determined by minimizing the residuals with respect to the physics equations and the observations. The resulting models are continuous in all dimensions, can zoom into local areas of interest in space and time, and provide information on physical parameters that are not necessarily directly observed such as horizontal velocities. Here we present the first proof of concept of this novel approach, explain the underlying methodology in detail, and provide an outlook to the many applications that PINNs enable.
\end{abstract}



\section{Introduction} \label{sec:introduction}
The understanding of observed astronomical phenomena greatly benefits from numerical models of the underlying phenomena. Observations of the Sun are frequently {\em inverted} to retrieve the physical properties of the solar atmosphere \citep[e.g.][]{DelToroIniesta2016}. Inversions generally deliver one-dimensional model atmospheres and pay great attention to the radiative transfer. The underlying hydrodynamics or magnetohydrodynamics equations are typically not used to constrain the retrieved physical quantities, although approximations to some of the equations are used in some of the most advanced inversion codes \citep[e.g.][]{Borrero2019}. Numerical simulations, on the other hand, perform ab initio simulations by calculating a solution of the radiative hydrodynamics equations as a function of time \citep[e.g.][]{Leenaarts2020}. The resulting time-dependent, three-dimensional models can only be compared with observations in a statistical sense since observational data cannot easily be included as a boundary condition in most numerical schemes.

Physics-Informed Neural Networks (PINNs) \citep{raissi2017physicsI, raissi2019physics, Karniadakis2021} offer a new approach to solving partial differential equations (PDEs) while matching observational data. PINNs solve PDEs by approximating the unknown solution with a deep neural network (DNN) that is trained to obey the PDEs. Data such as observations can easily be added as additional constraints when training the network. DNNs are particularly attractive when dealing with PDEs since derivatives of a DNN can be calculated analytically with little effort using automatic differentiation \citep[e.g.][]{Rumelhart1986}.

PINNs combine the advantages of inversions and numerical simulations while avoiding their respective weaknesses. Like numerical simulations, PINNs obey the time-dependent PDEs in a three-dimensional space and can include all relevant physics; like inversions, data can easily be assimilated by formulating them as additional constraints. PINNs also take full advantage of the enormous improvements that have been made in hardware and software for machine learning and artificial intelligence, and the precision can largely be selected by how long one trains the PINN. 

This paper serves as a proof of concept for using PINNs to generate physically self-consistent models that explain solar observations. Obviously, the more observations are included as constraints, the better the accuracy of the retrieved physical parameters at heights in the solar atmosphere from where the observed light originates. To proof the possibilities of the proposed approach, we chose the most simple and at the same time the most demanding application of PINNs: derive a physically consistent, time-dependent, three-dimensional model of the solar photosphere using only a sequence of continuum images at a single wavelength. As a first proof of principle many aspects are simplified and will need to be improved in the future. To assess the potential and limitations of the PINN approach, all developments and validations were done with established numerical simulations and synthetic images derived from these models. Applications to actual observations are beyond the scope of the present effort.

Section \ref{sec:methods} outlines the specific application of PINNs in detail, followed by Sect.\ref{sec:validation} where synthetic observations from an established numerical simulation are used to validate the approach and assess its capabilities in terms of retrieving physical parameters. We discuss the validation results and the limits of the current PINN implementation in Sect.\ref{sec:discussion} and close with an outlook to the many potential applications of PINNs in solar physics in Sect.\ref{sec:outlook}.

\section{Methods} \label{sec:methods}

\subsection{Numerical model} \label{subsec:bifrost}
For development and validation, we used results from Bifrost numerical simulations \citep{Gudiksen2011}. In particular, we used the {\tt qs024048\_by3363} quiet sun simulation available at \url{sdc.uio.no/search/simulations} and chose an area with minimal magnetic field. The horizontal resolution is 48~km and the vertical resolution around optical depth unity is about 20~km. We selected areas of 96 by 96 grid points horizontally and 64 grid points spanning -800~km to +500~km in height at the full resolution of the simulations, and 20 snapshots spaced by 50~s each. The four-dimensional model cube therefore spans 4600~km by 4600~km horizontally, 1300~km vertically, and 1000~s in time. For simplicity, we assume that the simulation cube is located at disk center where the gravity vector and the line of sight are parallel.

\subsection{Physics-Informed Neural Networks (PINNs)}
PINNs solve PDEs by approximating the unknown solution with a deep neural network that is trained to obey the PDEs over a domain $\Omega$, which, in our case, has a temporal and three spatial dimensions. The PDEs are therefore approximately solved by minimizing the PDEs reformulated as constraints. For instance, the continuity equation
\begin{equation}
  \frac{\partial\rho}{\partial t} = -\nabla\cdot\left(\rho\vec v\right)
\end{equation}
becomes
\begin{equation}
  \mathrm{min}\sum_\Omega\left|
    \frac{\partial\rho}{\partial t} + \nabla\cdot\left(\rho\vec v\right)
  \right|^2\;,
\end{equation}
where a deep neural network describes the density and three velocity vector components as a function of space and time. The weights of the deep neural network are then determined by minimizing the PDE constraints as well as the observational constraints.

Since PDE constraints are applied over the whole domain $\Omega$, it is crucial to formulate them in such a way that the {\em relative} residuals after minimization have the same order of magnitude across the whole domain and that the trivial solutions, such as a static solution for the continuity equation, are discouraged. We therefore calculate all PDE terms from the Bifrost simulation, plot them for all PDEs, and check that dominant terms do not vary more than a factor of a few over the full range of heights considered here. This is particularly important in the case of a stratified atmosphere where density and pressure drop off exponentially with height. In addition, these plots also reveal which terms are important at which heights. Furthermore, by formulating the PDE as constraints that should be equal to zero, we can multiply the constraints with arbitrary terms, which can greatly help with avoiding trivial solutions. Details for each PDE will be described in the sections below.

\subsection{Choice of physical parameters}
PINNs have the substantial advantage of letting us choose the physical parameters that the deep neural network should represent as a function of space and time. For the radiative hydrodynamics models presented here, the deep neural network determines the velocity vector components $v_x$, $v_y$, and $v_z$ in the $x$,$y$, and $z$-directions, the logarithm (base 10) of the density $\log\rho$, and the logarithm (base 10) of the temperature $\log T$. All physical quantities are in SI units. As density drops off exponentially with height, and the temperature also drops quickly with height, it is beneficial to use the logarithm of these two quantities for three reasons: 1) neural networks are not suited to represent quantities with large dynamical scales; 2) the (relative) residuals for density and temperature will be similar in magnitude at different heights in the atmosphere; and 3) the density and temperature themselves are guaranteed be positive.

Numerical simulations typically use the internal energy \citep[e.g.][]{Leenaarts2020} as a primary physical parameter since the energy equation provides a natural way to calculate the evolution of the internal energy as a function of time; the temperature and pressure then have to be determined by solving non-linear equations. Here we use the the logarithm of the temperature as the output of the PINN, which simplifies the calculation of the pressure and ionization fraction.

\subsection{Equation of State}
The momentum and energy PDEs require knowledge of the pressure. It can be calculated with the equation of state, which provides the pressure as a function of the density and the temperature. Here we only consider the ionization of hydrogen since that is of fundamental importance in convection \citep[e.g.][]{Rast1993}. Assuming that the hydrogen ionization is in equilibrium, we can calculate the ionization fraction of hydrogen, $X$, from the Saha equation
\begin{equation}
X^2/(1-X) =
s = \frac{1}{n}\left[\frac{2\pi m_{\rm e}kT}{h^2}\right]^\frac{3}{2}e^{\frac{-13.6{\rm eV}}{kT}}
  = \frac{5.53798845\cdot 10^{-6}}{\rho}T^{1.5} e^{-157821.45/T}\;,
\end{equation}
where $T$ is the temperature, $m_{\rm e}$ is the mass of the electron, $k$ is the Boltzmann constant, and $h$ is the Planck constant. $n$ is the number of neutral hydrogen particles per unit volume, which is calculated as
\begin{equation}
n = \frac{\rho}{1.29{\rm AMU} \cdot 0.934}\;,
\end{equation}
where we make the usual assumption that the mean molecular weight of the neutral solar-composition gas in the photosphere is 1.29 Atomic Mass Units and that the relative abundance of hydrogen atoms in the photosphere is 0.934 \citep[e.g.][]{Voegler2005}. To obtain $X$, we solve a quadratic equation in $s$. While this equation should always have a real-valued solution between zero and one, the neural network, particularly in the beginning, may be far away from a physically realistic scenario and return combinations of temperature and density that lead to complex solutions. We therefore take the absolute value to make sure that the calculation always succeeds and obtain the ionization fraction as
\begin{equation}
X = \frac{-s + \sqrt{\left|s^2 + 4s\right|}}{2}\;.
\end{equation}
Using the ideal gas law,
\begin{equation}
p = \frac{\rho}{1.29{\rm AMU}}(1+0.934X)kT\;,
\end{equation}
we can then calculate the natural logarithm of the total pressure from atoms and electrons as
\begin{equation}
\ln p = \ln\rho + \ln T  + \ln\left(1 + 0.934 X\right) + 8.7711096
\end{equation}
Here we use the natural logarithm of the pressure and density since the PDEs are most easily expressed in natural logarithms rather than base-10 logarithms. The DNN itself provides the base-10 logarithm of the density since that can be directly compared with the Bifrost simulations. We validated our pressure calculations by comparing to Bifrost simulations and found a better than 1\% agreement.

\subsection{Opacity and optical depth}
Since radiation is of fundamental importance in understanding solar granulation, we need to be able to calculate opacities and optical depths. To stay within the framework of analytically differentiable functions, we trained neural networks to return opacities as a function of temperature and pressure. The opacity networks consist of a scaling layer that normalizes the inputs (logarithm of pressure and logarithm of temperature) to the $\pm 1$ range, three fully connected, hidden layers with 4, 8 and again 8 nodes each and a hyperbolic tangent activation function followed by a fully connected single node and a linear activation function followed by a scaling layer that translates the output into physical units for the opacity per unit mass on a logarithmic scale. The network contains 133 free parameters and automatically interpolates between table values; derivatives are easily calculated with automatic differentiation.

\begin{figure}
  \includegraphics[width=0.48\textwidth]{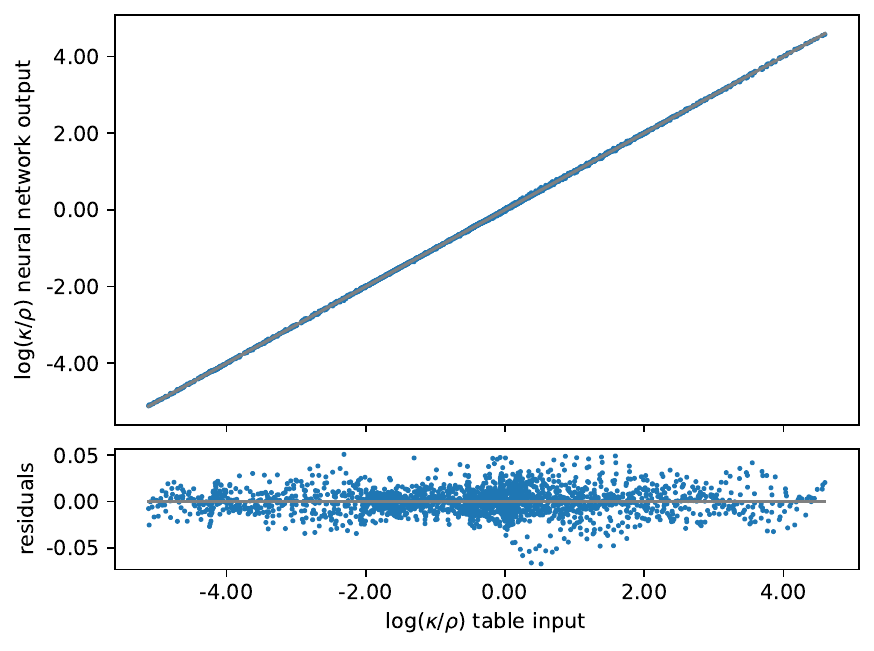}\hfill
  \includegraphics[width=0.48\textwidth]{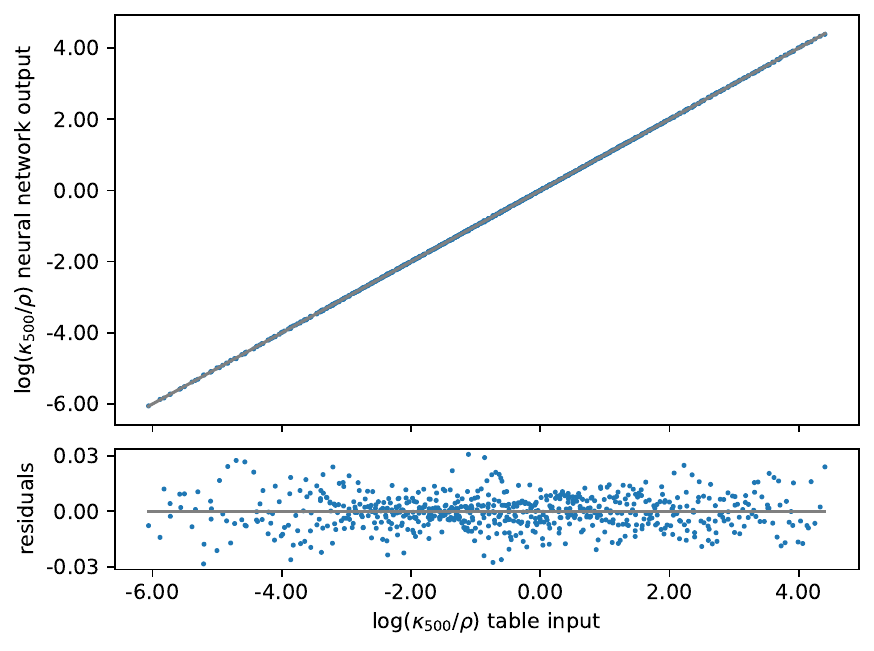}
  \caption{Rosseland mean (left) and continuum opacity at 500~nm (right) are calculated by neural networks that have been trained on tables. These neural networks with 133 parameters each reproduce the table values with high accuracy.}
  \label{fig:opann}
\end{figure}

Two versions of the opacity neural network have been trained: 1) Rosseland mean opacities to calculate optical losses and 2) continuum opacities at 500~nm to synthesize monochromatic continuum images. The Rosseland network was trained on the Rosseland mean opacity table {\tt kapp00.ros} by Fiorella Castelli (downloaded from \url{https://wwwuser.oats.inaf.it/fiorella.castelli/kaprossnew.html}) assuming zero turbulent velocity. Figure~\ref{fig:opann} shows the correspondence between table input and neural-network output. The residuals have an RMS of 0.014~dex. The continuum opacity network was trained on opacity tables created with the ABSKO package \citep{Gustafsson1973} for the relevant ranges of temperature and pressure.

The optical depths in the vertical direction are calculated with a numerical trapeze-rule integration of the opacity from the top down; this preserves the ability to use automatic differentiation of quantities related to optical depth but no longer provides the option to easily calculate the optical depth at arbitrary positions in space. \replaced{One could consider rewriting}{We also tried to rewrite} the optical-depth calculation as a PDE, which would then also be solved by the neural network. However, this made the optimization highly unstable and was therefore abandoned.

\subsubsection{Radiative loss}


\begin{figure}
  \includegraphics[width=\textwidth]{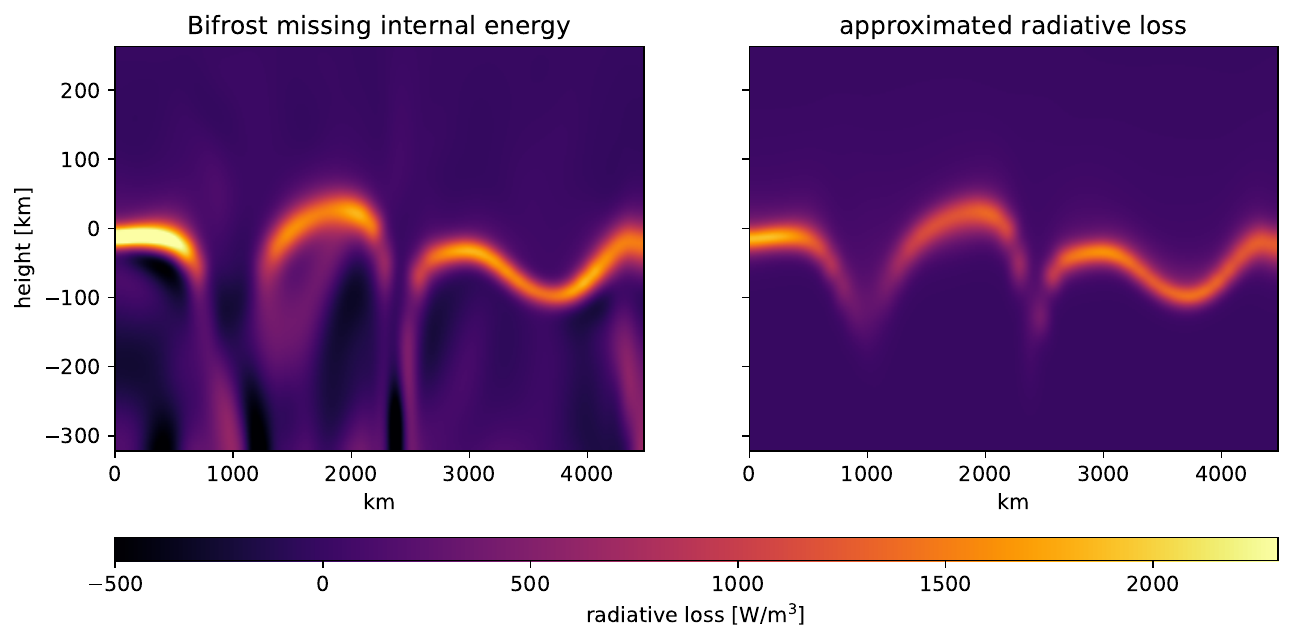}
  \caption{
    Comparison of missing internal energy in a Bifrost simulation and our radiative loss approximation, which folds horizontal losses into the vertical losses and can only account for radiative losses, not radiative heating as seen in the deeper layers of the Bifrost simulations.
  }
  \label{fig:qrad}
\end{figure}

Radiative loss calculations are computationally expensive. It would be natural to consider the diffusion approximation \citep{Mihalas1984} as its calculation only requires local quantities and their \added{first and second-order} derivatives, which are easy to calculate. \added{The radiative loss is the divergence of the radiative flux; in the diffusion approximation the radiative flux is proportional to the temperature gradient. As such, the diffusion approximation of the radiative loss requires the calculation of second-order derivatives of the temperature with respect to the spatial coordinates.} However, for the particular hardware and software used here, calculations of second-order derivatives significantly slowed down the calculations, and we therefore abandoned that approach.

For this proof-of-principle effort, we chose to use a highly simplified approach to the radiative loss calculation inspired by the approximation in \cite{AbbettFisher2012} where the integrations over angle and optical depth are carried out analytically. The second-order elliptical integral $E_2$ that is central in \cite{AbbettFisher2012} can further be approximated by $E_2(\tau)\approx e^{-\tau}/(1+\tau)$. We then used Bifrost simulations to fit three free parameters and ended up with the radiative loss per unit mass of  
\begin{equation}
q_{\rm rad} = 0.3366811 \kappa_R \sigma T^4 e^{-0.0178391\tau_R} / (1.0 + 4.440909\tau_R)\;,
\end{equation}
where $\kappa_R$ is the local Rosseland mean opacity per unit mass, $\sigma = 5.67\cdot 10^{-8}$ W/m$^2$/K$^4$ is the Stefan-Boltzmann constant, and $\tau_R$ is the corresponding optical depth in the vertical direction. To estimate the radiative loss in the Bifrost simulations, we used the energy equation (see \ref{sec:energy}) where we can calculate all the terms except for the radiative loss term. In addition to fitting the Bifrost simulations, we also required that the total radiative energy be close to the solar value of $6.3\cdot 10^7$ W/m$^2$. Note that we should have used the Planck-weighted mean opacity instead of the Rosseland opacity. However, in the layers of the photosphere that we are considering here, the Planck opacity is largely a scaled version of the Rosseland mean opacity \citep{Przybylski1960}, and that scaling is taken into account by fitting the parameters of the equation above.

Figure~\ref{fig:qrad} shows a comparison between the radiative loss in the Bifrost simulations and our radiative loss approximation. Our radiative loss approximation will always return positive values; therefore, it is unable to capture the radiative heating as seen in the Bifrost simulations as negative radiation losses. Since we constrain the vertically integrated radiative loss to be equal to the solar radiative output, our losses are, on average, smaller at the $\tau=1$ surface than in the Bifrost simulations. Furthermore, our approximation, by design, is not able to reproduce \replaced{the non-vertical radiation losses seen in the Bifrost simulations}{radiative losses in areas where a locally plane parallel atmosphere is an inadequate approximation such as in areas where large horizontal temperature gradients occur}. However, it does compensate for the lack of \replaced{non-vertical}{these horizontal} losses by increasing the \replaced{vertical component}{radiative loss above those areas} as can be seen above intergranular lanes where our approximation exhibits larger \deleted{vertical} losses than the Bifrost simulations. In the future, the radiative losses surely need to be calculated much more accurately by taking into account the angle and wavelength dependence of the losses. For now, the simple approximation seems to be good enough to prove the principle of using PINNs to retrieve atmospheric parameters from image sequences.

\subsection{Continuity equation} \label{subsec:continuity}

\begin{figure}
  \includegraphics[width=0.48\textwidth]{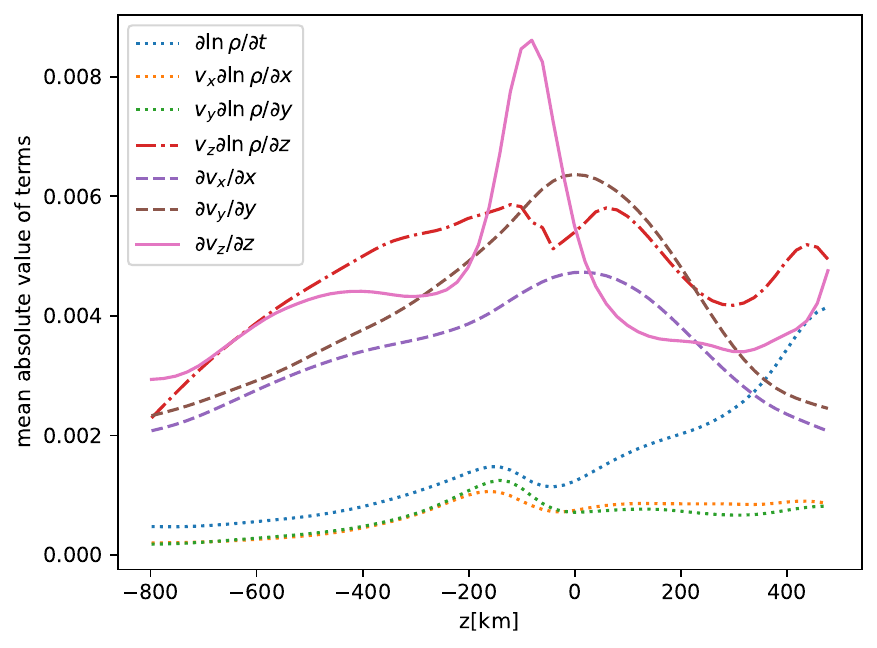}
  \caption{Horizontally averaged absolute magnitudes of all terms of the continuity equation as a function of height in a Bifrost simulation of the quiet Sun.}
  \label{fig:cont}
\end{figure}

The conservation of mass requires the continuity equation to be fulfilled everywhere, and we therefore minimize the following constraint:
\begin{equation}
  \frac{1}{\sigma^2_{v_z}}
  \sum_\Omega\left[
    \frac{\partial\ln\rho}{\partial t} + 
    v_x\frac{\partial\ln\rho}{\partial x} + v_y\frac{\partial\ln\rho}{\partial y} + v_z\frac{\partial\ln\rho}{\partial z} + 
    \frac{\partial v_x}{\partial x} + \frac{\partial v_y}{\partial y} + \frac{\partial v_z}{\partial z}\right]^2\;,
\end{equation}
where $\sigma^2_{\rm v_z}$ is the variance of the vertical velocity. Since the term $\partial\ln\rho / \partial t$ tends to be small compared to the other terms except for the uppermost photosphere (see Fig.~\ref{fig:cont}), the easiest way to fulfill the continuity equation is to minimize the velocities. To avoid this trivial solution, we divide by the variance of the vertical velocity. The vertical velocity is fundamental for the convective energy transport and will therefore be estimated quite accurately from the observations. The horizontal velocities, on the other hand, are largely deduced from obeying the continuity equation. We therefore chose to only use the vertical velocity to normalize the continuity constraint.

The magnitude of the most important terms in the continuity equation vary in height but the variation is typically less than an order of magnitude. The dominant terms peak around a height of zero, corresponding to optical depth unity. This implies that the minimization of the continuity constraint will lead to smaller relative residuals around optical depth unity where most of the observed light originates from. The resulting model will therefore enforce the continuity equation the strongest where the observations contain information about the atmosphere, which is a beneficial side effect.

\subsection{Momentum equations}\label{subsec:momentum}

\begin{figure}
  \includegraphics[width=0.48\textwidth]{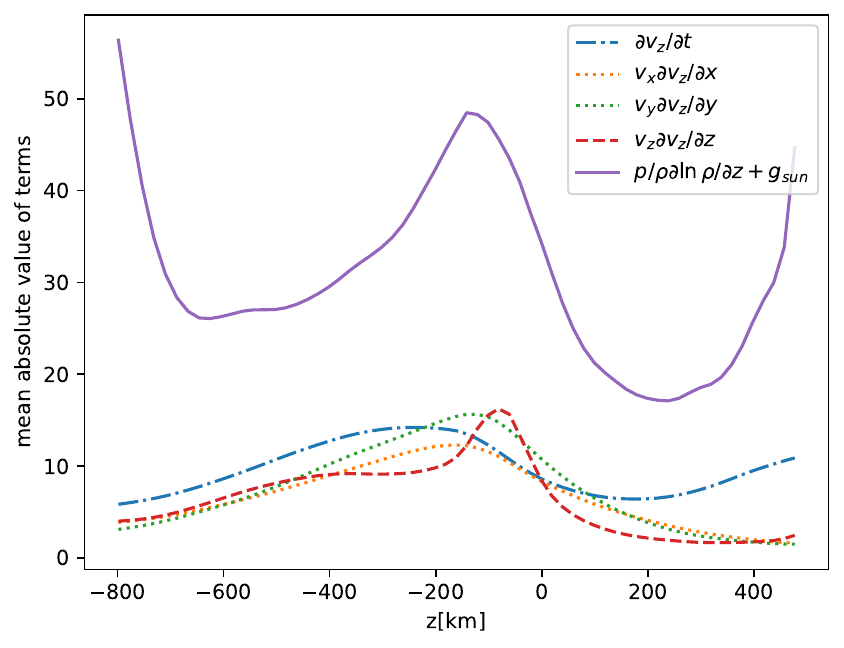}
  \caption{Horizontally averaged absolute magnitudes of all terms of the vertical momentum equation as a function of height in a Bifrost simulation of the quiet Sun.}
  \label{fig:momv}
\end{figure}

\begin{figure}
  \includegraphics[width=0.48\textwidth]{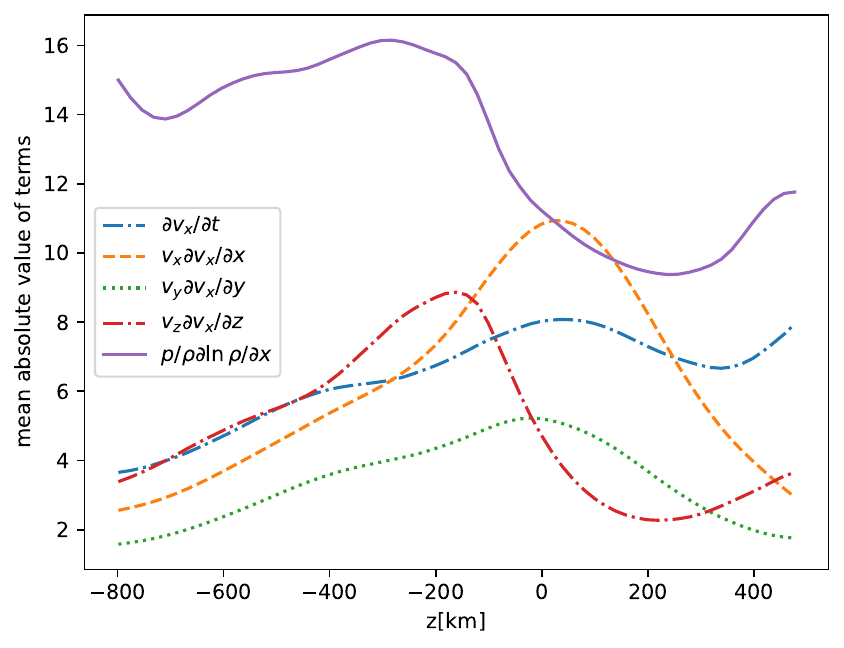}\hfill
  \includegraphics[width=0.48\textwidth]{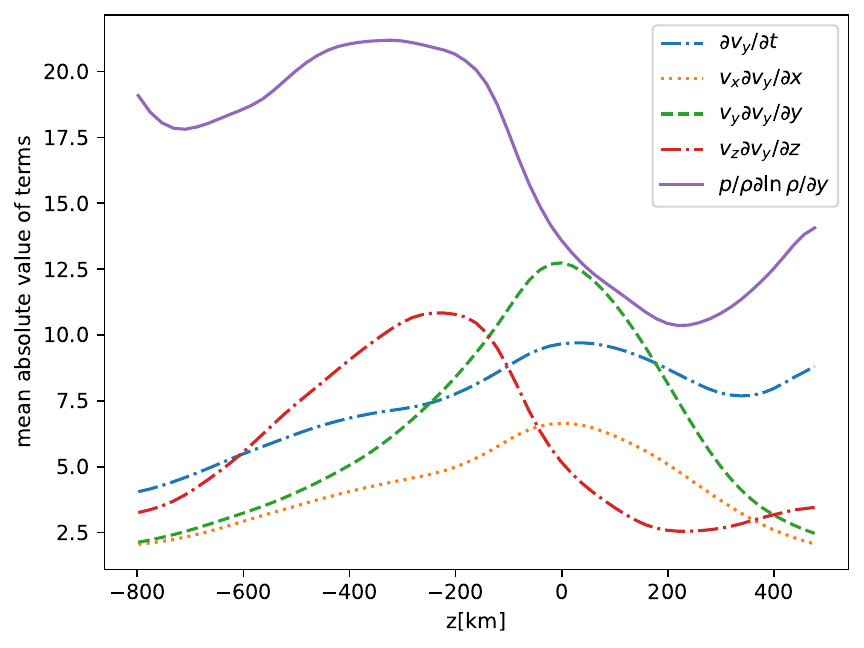}
  \caption{Horizontally averaged absolute magnitudes of all terms of the horizontal momentum equations as a function of height in a Bifrost simulation of the quiet Sun.}
  \label{fig:momh}
\end{figure}

The momentum PDEs in the $x,y$, and $z$ axes, formulated as constraints, are
\begin{equation}
\sum_\Omega\left[
  \frac{\partial v_x}{\partial t} + v_x\frac{\partial v_x}{\partial x} + v_y\frac{\partial v_x}{\partial y} + v_z\frac{\partial v_x}{\partial z} + 
  \frac{p}{\rho} \frac{\partial \ln p}{\partial x}
\right]^2\;,
\end{equation}
\begin{equation}
  \sum_\Omega\left[
    \frac{\partial v_y}{\partial t} + v_x\frac{\partial v_y}{\partial x} + v_y\frac{\partial v_y}{\partial y} + v_z\frac{\partial v_y}{\partial z} + 
    \frac{p}{\rho} \frac{\partial \ln p}{\partial y}
  \right]^2\;,
\end{equation}
and

\begin{equation}
  \frac{1}{\sigma^2_{v_z}}\sum_\Omega\left[
    \frac{\partial v_z}{\partial t} + v_x\frac{\partial v_z}{\partial x} + v_y\frac{\partial v_z}{\partial y} + v_z\frac{\partial v_z}{\partial z} + 
    \frac{p}{\rho} \frac{\partial \ln p}{\partial z} + g_{\rm sun}
  \right]^2\;,
\end{equation}
where $g_{\rm sun}$ is the solar gravitational force.

For the momentum equation in the $z$-direction we found it again useful to normalize with the variance of the velocity in the $z$-direction. The gravitational force and the vertical pressure gradient are the two dominant terms and largely compensate each other. Similar to the continuity equation, a vertically static atmosphere with $v_z=0$ is therefore again a simple approach to fulfill the vertical momentum equation, a solution that is obviously not correct.

In contrast, the momentum equations in the $x$ and $y$ direction are not normalized. The \replaced{continuum}{continuity} equation is largely sensitive to the spatial derivatives of the horizontal velocities, and not the velocities themselves. By not normalizing the momentum equations in the horizontal directions, we suppress unrealistically large horizontal velocities.

\subsection{Internal energy equation} \label{sec:energy}
The internal energy equation is expressed in terms of the internal energy per unit mass, $e$, to avoid the roughly exponential decrease with height that the internal energy itself exhibits. The internal energy of the gas per unit mass is given by
\begin{equation}
  e = \frac{3p}{2\rho} + \frac{0.934 X \cdot 13.6{\rm eV}}{1.29{\rm AMU}}\;,
\end{equation}
where the first term corresponds to the internal energy per mass of an ideal gas and the second term describes the hydrogen ionization energy per mass \citep[e.g.][]{Voegler2005}.

\begin{figure}
  \includegraphics[width=0.48\textwidth]{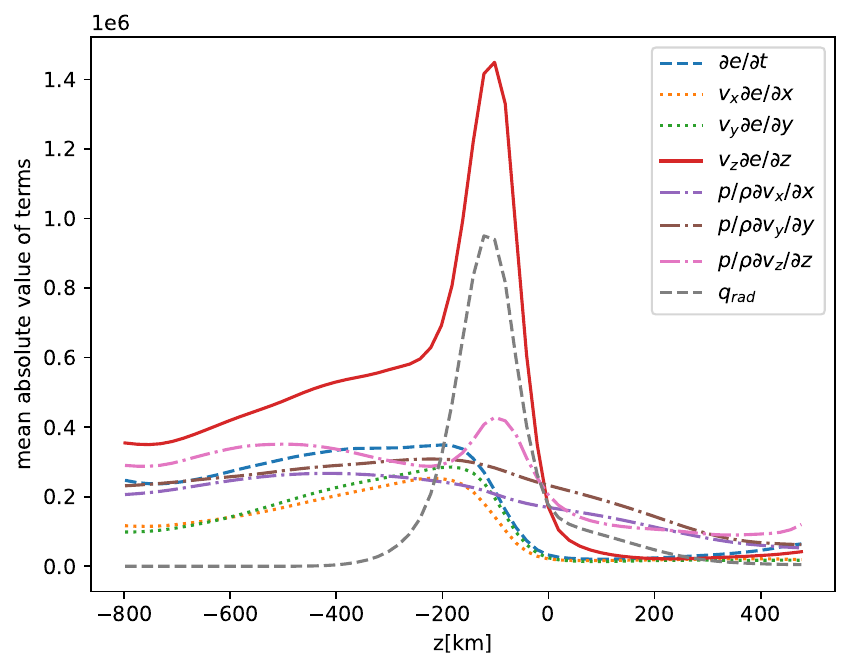}
  \caption{Horizontally averaged absolute magnitudes of all terms of the energy equation as a function of height in a Bifrost simulation of the quiet Sun.}
  \label{fig:ener}
\end{figure}

The internal energy PDE constraint is then given by
\begin{equation}
\sum_\Omega\left[\frac{\partial e}{\partial t} + 
  v_x\frac{\partial e}{\partial x} + v_y\frac{\partial e}{\partial y} + v_z\frac{\partial e}{\partial z} + 
  \frac{p}{\rho}\left(\frac{\partial v_x}{\partial x} + \frac{\partial v_y}{\partial y} + \frac{\partial v_z}{\partial z}\right) + 
  q_{\rm rad}
\right]^2\;.
\end{equation}
Note that this does not include viscous and conductive heating, which are small in the photospheric layers (M.~Rempel, private communication). \deleted{Since kinetic and acoustic energy fluxes largely cancel at depths that are considered here \citep{SteinNordlund1998}, it should be sufficient to only consider the evolution of the internal energy.}

\subsection{Additional constraints} \label{subsec:constraints}
The continuity and momentum equations constrain the PINN locally; the energy equation, except for the radiative loss term, also only depends on local quantities. The simulated observations are obtained by integrating the radiative transfer equation in the vertical direction where meaningful contributions only come from a relatively small range in height around optical depth unity. To obtain realistic results at depths and heights where no direct observational constraints are available, it therefore greatly helps to add global constraints that act on the whole volume. The more global constraints we can add, the more realistic the model will become, particularly in layers where observations do not provide direct constraints. 

Since all observations are linked to optical depth and not geometrical depth, it is possible to translate a model in the $z$-direction without affecting any of the observational parameters or PDE constraints. To discourage the model from arbitrary translations in the $z$-direction, we fix the average density $\log\rho_{\rm sun}$ and temperature $\log T_{\rm sun}$ stratifications at every time step. This also helps to quickly achieve reasonable stratifications as the PINN at the start of the optimization is unaware of any of the physics that is involved.

Constraining the average density and temperature stratifications are implemented as constraints that have to be minimized at all heights and at all times:
\begin{equation}
  \frac{1}{N_z N_t}\sum_{z,t}\left[
    \log\rho_{\rm sun}(z) - \frac{1}{N_x N_y}\sum_{x,y}{\log\rho(x,y,z,t)}
  \right]^2
\end{equation}

and

\begin{equation}
  \frac{1}{N_z N_t}\sum_{z,t}\left[
    \log T_{\rm sun}(z) - \frac{1}{N_x N_y}\sum_{x,y}{\log T(x,y,z,t)}\;.
  \right]^2
\end{equation}

In addition, we require that the net mass flow through any $(x,y)$-plane at any time or any height are minimized, which is implemented by
%
\begin{equation}
  \frac{1}{N_z N_t}\sum_{z,t} \left[\frac{
    \sum_{x,y}{\rho v_z}}{
    \sum_{x,y}\rho}
  \right]^2\;.
\end{equation}
This prevents any spurious global waves from building up and makes sure that any mass flows going up are compensated by similar mass flows going downwards. The latter is particularly important to obtain realistic flow patterns in the deeper layers.

Similarly we require the total energy flux to be the same at all heights and times, which is implemented with the following constraint
\begin{equation}
  \frac{1}{N_z N_t}\sum_{z,t}\left[
    \frac{1}{N_x N_y}\sum_{x,y}{\left(\rho v_z - \left<\rho v_z\right>\right)
    \left(e+\frac{p}{\rho}+\frac{1}{2}v^2 \right) + q_{\rm rad}\rho} -
    6.3\cdot 10^7
  \right]^2  \;,
\end{equation}
where $6.3\cdot 10^7$ W/m$^2$ is the average energy per square meter emitted by the solar photosphere \citep[e.g.][]{Leenaarts2020} and $\left(\rho v_z - \left<\rho v_z\right>\right)$ is the net vertical mass flux \citep{SteinNordlund1998}, which is calculated by removing the horizontally averaged mass flow. The latter should approach zero due to the net mass flow constraint above, but it turned out to be difficult to make these two constraints work in harmony when the energy-flux constraint assumes that the net-mass-flux constraint is already fulfilled. \added{Since kinetic and acoustic energy fluxes largely cancel at depths that are considered here \citep{SteinNordlund1998}, we ignored their contributions to the total energy flux.}

\subsection{Observational constraint}
For this proof of concept, we limit the observational constraints to monochromatic continuum images at 500nm. The intensity is obtained \replaced{via a solution to}{from the PINN output by numerically integrating} the \deleted{formal} equation of radiative transfer in local thermal equilibrium and in the vertical direction\replaced{,}{. The observational constraint is then implemented as the sum of squares over the difference between observed images and the calculated continuum intensities,} 
\begin{equation}
  \sum_{x,y,t}\left[I_{\rm obs}(t) -
  \frac{2hc^2}{\lambda^5}\int\frac{e^{-\tau}}{e^\frac{hc}{\lambda kT}-1}d\tau
  \right]^2\;,
\end{equation}
where the continuum optical depth $\tau$ is calculated at 500~nm and $I_{\rm obs}(t)$ are the observed images as a function of time. The integral over optical depth is implemented via a numerical scheme employing the trapezoid rule and summing from the bottom up.

\added{In the following we will not use real observations but synthetic continuum images created from the deep neural network that was directly trained on the Bifrost model (see \ref{subsec:dnn} below) and provides an excellent representation of the Bifrost data in exactly the same format as the PINN. As such, the synthetic observations are calculated in exactly the same way as described in the previous paragraph for the PINN.}

\subsection{Physics sampling}
While the PINN itself is continuous in space and time, the constraints have to be calculated at discrete points in the domain $\Omega$. For now, we use a regular grid in all four dimensions, although that is in no way a requirement for the PINN approach to work. Since the computational effort scales with the number of grid points, it is advantageous to first train the neural network on a coarser grid before going to a fine grid to recover all the details. For instance, by decreasing the number of grid points by a factor of two in all four dimensions, the calculations can be sped up by about a factor of $2^4=16$.

Currently, the grid spacing in time $t$ and horizontal directions $x$ and $y$ is determined by the images. Ultimately, the grid spacing needs to be determined by the physics described by the PDEs. While classical time-integration schemes have well-understood requirements for the time step \citep{Courant1928} as well as numerical viscosities and diffusivities \citep[e.g.][]{Voegler2005}, these aspects are not yet well understood in PINNs. One might worry that the PINN solution may greatly violate the physics constraints between points where the constraints are calculated; this is actually unlikely to happen because of the spectral bias of deep neural networks \citep{Rahaman2018}, which describes the tendency of neural networks to fit the large scale much better than the small scales. As such, the spectral bias dampens small-scale fluctuations and acts as an artificial diffusivity that suppresses small-scale fluctuations. Note that the spectral bias cannot be remedied by just increasing the grid spacing where the constraints are enforced. 

\begin{figure}
  \includegraphics[width=0.48\textwidth]{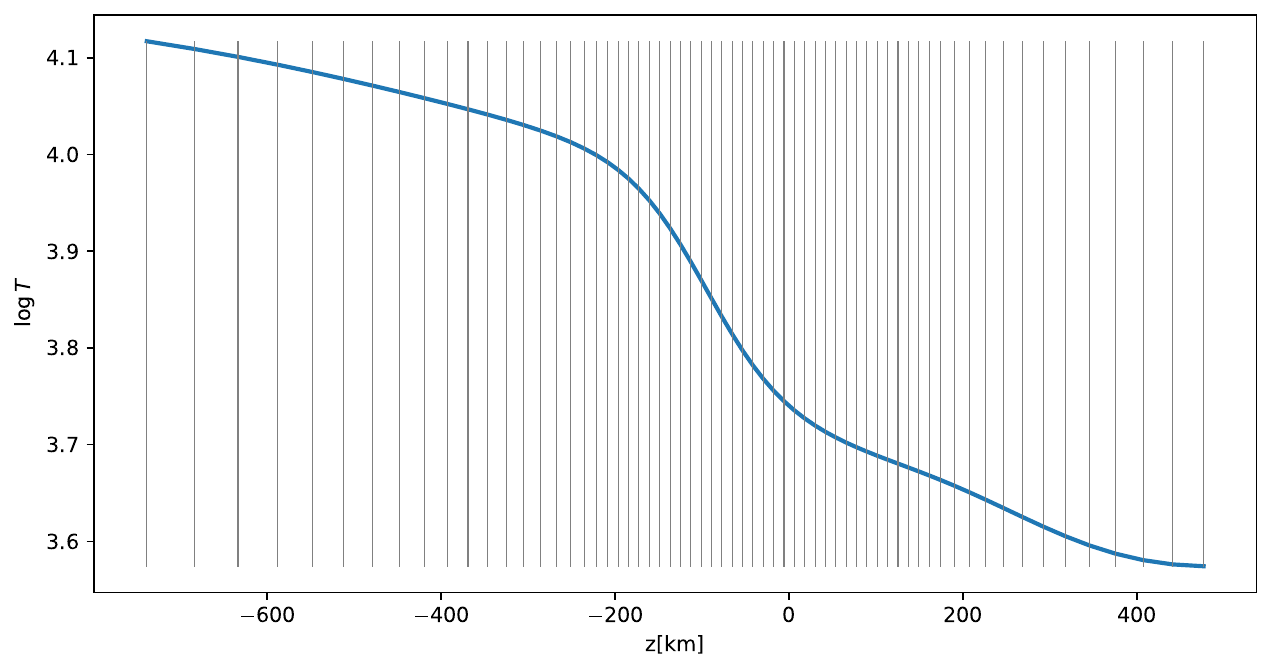}
  \caption{The logarithm of the mean temperature in the Bifrost simulation as a function of height. The vertical grey lines indicate the locations along the $z$-axis where the physics constraints are evaluated. The sampling is denser where the energy transitions from convective to radiative energy and where the physical parameters most quickly change with height.}
  \label{fig:zaxis}
\end{figure}

Close to the optical depth $\tau=1$ where convective energy is transformed into radiative energy, the physical quantities change rapidly. We therefore use an unevenly sampled grid in the $z$-axis (see Fig.~\ref{fig:zaxis}). This reduces the number of locations in $z$ where the PDEs have to be constrained while providing sufficiently dense samples to enable accurate calculations of synthetic images.

\subsection{Deep Neural Network design}\label{subsec:dnn}
The current implementation of the deep neural network was determined by trial and error. The deep neural network consists of a rescaling layer to bring the physical coordinates $x,y,z,t$ into the $\pm 1$ range, which is where neural networks work best, followed by five hidden, fully connected layers with 32, 64, 128, 128 and 128 nodes, respectively, each using the hyperbolic tangent as an activation function. This is followed by an output layer with five nodes and a linear activation function. The last layer scales the neural-network output back to physical units for the velocity vector and the logarithms of the density and the temperature. The network contains a total of 44,261 free parameters.

\begin{figure}
  \includegraphics[width=\textwidth]{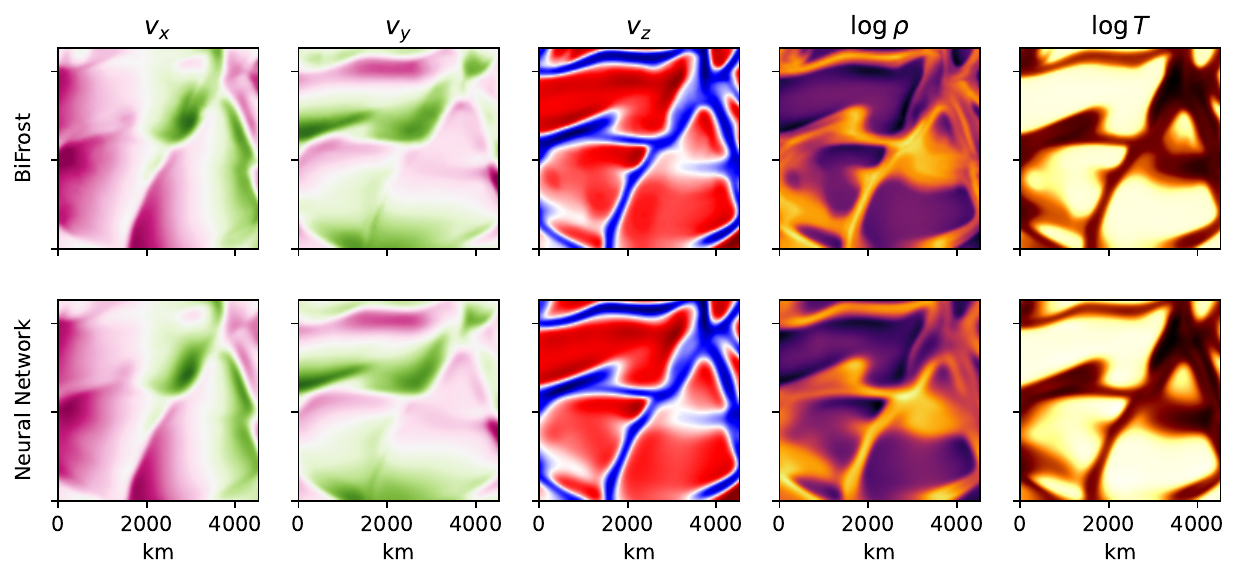}
  \caption{Comparison of horizontal slices \added{(in the middle of the the time sequence and at $40km$ below the average $\tau=1$ height in the Bifrost simulations where the vertical velocity gradient reaches its maximum)} through the parameters of the Bifrost simulation (top) and the output of a deep neural network (bottom) that has been trained to directly reproduce the Bifrost time series. The deep neural network has the same structure as the PINN but is not informed by any physics constraints. Apart from a slight loss in spatial resolution, the neural network reproduces the numerical simulations to a high degree. This indicates that the chosen deep neural network architecture is able to closely reproduce the Bifrost simulations except for the smallest details.}
  \label{fig:bifnn}
\end{figure}

\begin{figure}
  \includegraphics[width=0.48\textwidth]{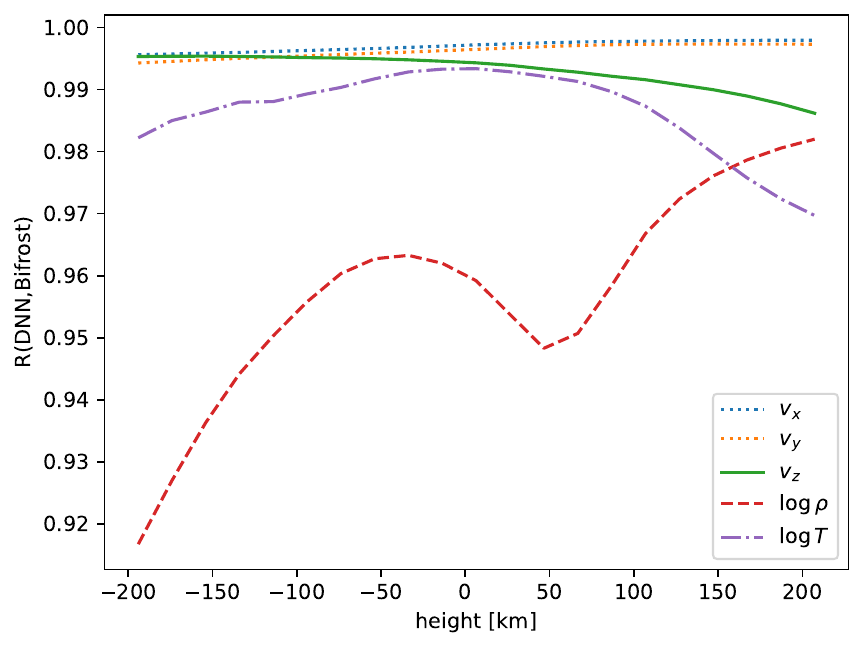}
  \caption{Pearson correlation coefficients as a function of height between the Bifrost model and the deep neural network for the three velocity components and the logarithms of the density and the temperature. The height scale has been adjusted so that $z=0$ corresponds to the average height were the optical density at 500~nm reaches unity. The correlation coefficient is high, indicating that the chosen deep neural network architecture is able to closely reproduce the Bifrost simulations.}
  \label{fig:corrbif}
\end{figure}

The main requirement for the network is to be able to adequately represent the Bifrost simulations themselves. If it cannot represent the simulations with sufficient accuracy, then the PINN will never be able to represent a solution that is close to the Bifrost simulations. We therefore trained a network with this structure directly on the Bifrost simulations and determined the deviations between the neural-network output and the Bifrost simulations. Fig.~\ref{fig:bifnn} shows a visual comparison of the Bifrost simulations and the trained deep neural network at a height close to where the average optical depth is unity. Apart from a slight loss in spatial resolution, which is most noticeable in the density, the deep neural network is able to represent the Bifrost simulations quite well. Indeed, the mean absolute error of the neural-network output of all five physical parameters is about 1\% of the full range of the respective parameter. Figure~\ref{fig:corrbif} shows the high degree of correlation between the deep neural network and the Bifrost model that it was trained on. The density shows a markedly lower correlation coefficient because the density shows structures at smaller scales than any of the other parameters; the inability of the deep neural network to reproduce the smallest-scale features is therefore most pronounced in the density parameter.

The deep neural network representation of the Bifrost simulations is highly compressed compared to the original Bifrost data: the deep neural network has 44,261 floating point values whereas the Bifrost simulation cube has $96\cdot 96\cdot 64\cdot 20 = 11,796,480$ floating point values. This corresponds to a compression by a factor of 267; this compression factor may be different for other numerical simulations and depends on their smoothness. In addition, the deep neural network can be evaluated at any point in space and time covered by the simulations, therefore enabling a trivial way to interpolate between grid points; furthermore, gradients are calculate with high efficiency with automatic differentiation. As such, it might be worth considering releasing deep-neural-network representations of (magneto)hydrodynamic simulations.

\subsection{Starting values}
Training a neural network is simply a huge, non-linear optimization problem. As for any optimization effort, a good starting point for the solution is essential for a fast convergence. As such, we developed a neural network, {\tt int2mod} that can estimate the vertical stratification of the density, the temperature and the vertical velocity based on the normalized (mean = 0; rms = 1) continuum intensity. This simple approach is motivated by the fact that there is a strong correlation between intensity and vertical velocity, and in turn also between velocity, temperature, and density \citep{SteinNordlund1998}. Since the instantaneous, local continuum intensity only contains information from a shallow layer, it is crucial to add additional constraints to produce somewhat realistic stratifications in the layers below and above where the continuum radiation emerges from. Therefore, we also constrain the neural-network output to preserve some of the statistics and correlations between the different physical parameters: zero net mass motion through each horizontal layer, mean and variance of the density and temperature stratifications (see Sect.~\ref{subsec:constraints}), as well as the variance of the vertical velocity and the correlation between mass flux and temperature, both as a function of height. The network is trained on a part of the Bifrost simulation that is not used otherwise.

The {\tt int2mod} neural network takes the normalized intensity and the $z$ coordinate as inputs and returns the vertical velocity and the logarithms of the density and the temperature as output. The neural network has 811 free parameters and consists of a rescaling layer to bring the $z$ coordinate into the $\pm 1$ range (the intensity is already normalized) followed by three hidden, fully connected layers with 8, 16, and 32 nodes, respectively, each using the hyperbolic tangent as an activation function. This is followed by an output layer with three nodes and a linear activation function. The last layer scales the neural-network output back to physical units. In the future, one might switch to much more sophisticated convolutional neural networks along the lines of DeepVel \citep{AsensioRamos2017} or even networks trained on large-scale MHD simulations \citep{Yang2024}.

The {\tt int2mod} neural network is then used to generate the starting values by first translating an intensity image into four-dimensional cubes of the vertical velocity and the logarithms of the density and temperature on an $y,x,t,z$ grid; the reason for this particular order of axes is explained in Sect.~\ref{subsec:implementation}. To obtain an initial version of the PINN network, we train the PINN on the output of the {\tt int2mod} network. In addition, we also estimate the horizontal velocities from the continuity equation, or in this case its anelastic approximation, formulated as a constraint

\begin{equation}
  \sum_\Omega\left[
    v_z\frac{\partial\ln\rho}{\partial z} + 
    \frac{\partial v_x}{\partial x} + \frac{\partial v_y}{\partial y} + \frac{\partial v_z}{\partial z}\right]^2\;.
\end{equation}

The anelastic approximation is appropriate everywhere except for the higher photospheric layers (see Fig.~\ref{fig:cont}). Since the anelastic approximation is insensitive to an arbitrary constant horizontal velocity offset, we penalize the mean square horizontal velocity to avoid unrealistically large horizontal velocities. In addition, we only use the anelastic approximation of the continuity equation to determine the horizontal velocity components $v_x$ and $v_y$; the density and vertical velocity are fitted to the {\tt int2mod} results and are not influenced by the continuity equation to avoid a back-reaction from the horizontal velocities on the density and vertical velocity estimates.

\subsection{Loss scaling} \label{subsec:loss}
The merit function that is being minimized, or loss function as it is called in machine learning, consists of 10 different constraints: observed image, continuity equation, momentum equations in $x$, $y$, and $z$, energy equation, mass flux constraint, energy flux constraint, and mean stratifications in density and temperature. Each constraint is multiplied with a scalar loss scale before they are all added up as the total loss function. These loss scales were determined such that the loss terms were all of the same order of magnitude when using the deep neural network that was fitted directly to the Bifrost simulation. In addition, we checked that when using this directly fitted deep neural network as a starting point for an iterative optimization that the solution did not significantly deviate from the starting point. During development, the latter test also quickly revealed when errors were present in the code.

\subsection{Implementation} \label{subsec:implementation}
The code \replaced{as}{\citep{Keller_2025_15346957} is} available at \url{https://github.com/cukeller/rhpinn.git}\added{,} is based on TensorFlow 2.16.1 and Keras 3 and was tested on a MacBook Pro with an M3Max Apple Silicon processor with a 40-core GPU and 64GB of RAM. The calculations take at most 10 minutes. We use this hardware and software combination to efficiently solve our large-scale optimization problem. We use the Adam optimizer \citep{KingmaBa2015} and tuned the learning rate, the batch size and the number of epochs by hand to maximize the convergence speed; the choice of these parameters depends on the hardware that is being used. As the optimization does not necessarily lead to a monotonically decreasing loss function, we saved all free parameters whenever the loss function reached a new minimum. 

A crucial aspect of the code is the order of the different axes. Here we chose the order, from slowest to fastest varying index, as $y$, $x$, $t$, $z$. This is largely driven by how machine-learning optimizers work. Not all data are fitted at once, but typically in batches. Since comparisons with observations require radiative transfer calculations in the $z$-direction, it is crucial that all $z$-positions for a given $x$ and $y$ are included in a single batch. Furthermore, quantities that need to be conserved at all times, as compared to an average over time, require that each batch also contains every time step; this is particularly important for the additional constraints in Sect.~\ref{subsec:constraints}. For each epoch, we therefore shuffle with respect to all $x$ and $y$ positions but leave the order of points in the $t$ and $z$-directions untouched to allow easy integration in the $z$-direction. 

In general, we start with calculating the constraints only on every other grid point in $x$, $y$, and $t$ to speed up the optimization by about an order of magnitude; we do not reduce the number of grid points in the $z$-direction as that would reduce the quality of the radiative transfer calculations. Once converged, we continue iterating with the constraints being evaluated on every grid point to recover smaller details.

\section{Validation} \label{sec:validation}

Validation of our PINN approach is crucial as it is not obvious that it actually works; it could easily happen that the optimization is trapped in a local minimum that is far away from the global minimum. Indeed, much of the development effort consisted of choosing and scaling the constraints in such a way as to avoid instabilities in the optimizer and being trapped in local minima far from a reasonable solution.

\begin{figure}
  \includegraphics[width=\textwidth]{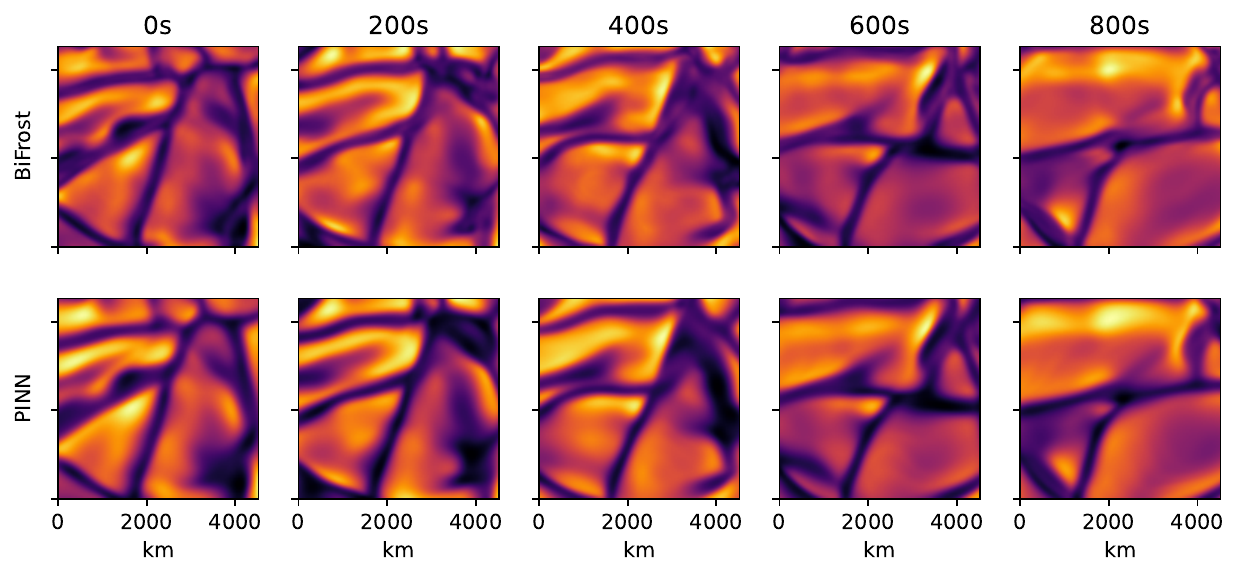}
  \caption{Comparison of continuum images at 500~nm calculated from Bifrost simulations (top row) and the corresponding images calculated from the PINN model. The images span a period of 800 seconds. As expected, the PINN images show slightly less fine structure as compared to the Bifrost images.}
  \label{fig:images}
\end{figure}

For the validation, we created a sequence of 20 continuum images at $\lambda=500$~nm from the Bifrost simulations whose properties were described in Sect.~\ref{subsec:bifrost}. The emerging intensity $I_{500}$ was calculated with the trapezoid scheme applied to
\begin{equation}
  I_{500} =
  \frac{2hc^2}{\lambda^5}\int\frac{e^{-\tau_{500}}}{e^\frac{hc}{\lambda kT(\tau_{500})}-1}d\tau_{500}\;,
\end{equation}
where $\tau_{500}$ is the optical depth at 500~nm. These 20 images were then used as constraints for training the PINN. Figure~\ref{fig:images} compares five input images and the corresponding images calculated from the atmospheric model represented by the PINN. The correspondence is excellent, which is not too surprising as the input images constrain the PINN itself.

\begin{figure}
  \includegraphics[width=\textwidth]{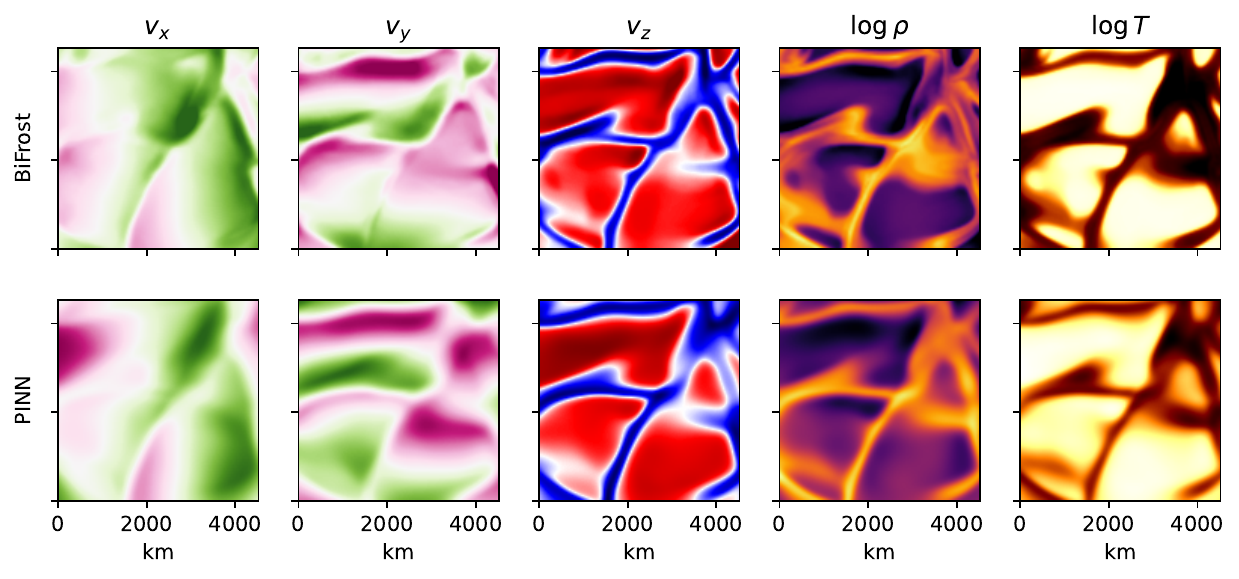}
  \caption{Comparison of physical parameters in the middle of the time sequence from a horizontal slice (same height as in Fig.~\ref{fig:bifnn}) through the Bifrost simulations (top row) at a height from which most of the continuum intensity emerges and the corresponding physical parameters from the PINN model. The PINN model was trained only on the continuum images calculated from the corresponding Bifrost simulation.}
  \label{fig:bifpinn}
\end{figure}

A much more demanding validation comes from comparing the physical parameters themselves. Figure~\ref{fig:bifpinn} shows horizontal slices in the middle of the time series from a height where most of the radiation emerges. There is generally good agreement between the Bifrost simulation and the PINN model, showing that the PINN approach works. This holds true for the whole sequence, and there is no noticeable degradation at the beginning or end of the time sequence. While the signs of the horizontal velocities are recovered well, their amplitudes show considerable discrepancies. 

\begin{figure}
  \includegraphics[width=\textwidth]{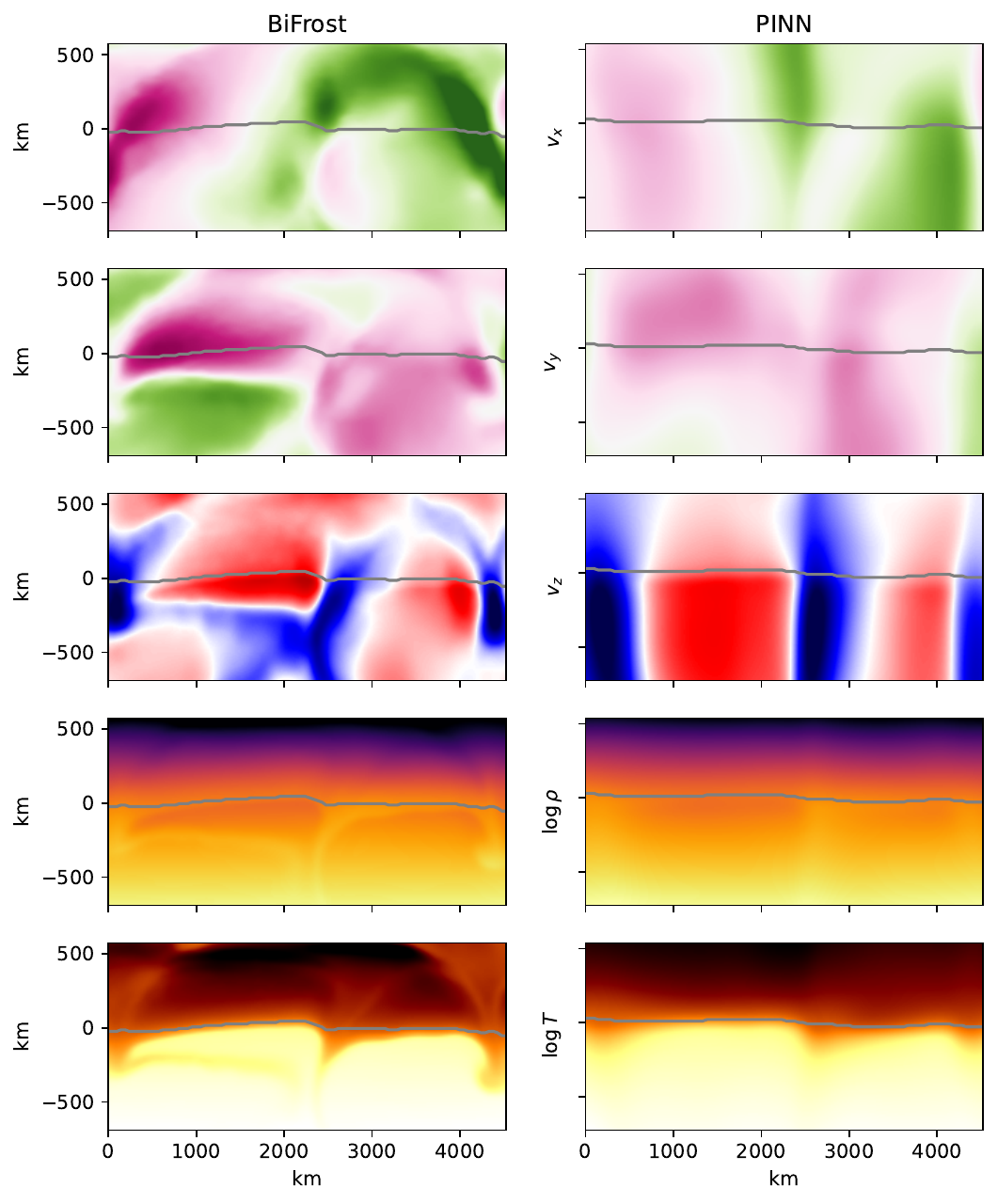}
  \caption{Same as Fig.\ref{fig:bifpinn} but for a vertical slice. The grey lines indicate the optical depth $\tau=1$ at 500~nm in the respective models; the vertical axes have been adjusted separately for the two models to make the average optical depth unity correspond to $z=0$.  There is decent agreement between the Bifrost simulation and the PINN model, with the best agreement around $\tau=1$ where most of the continuum intensity forms and worse agreement in layers where the continuum images do not contain any relevant information.}
  \label{fig:bifpinnv}
\end{figure}

Figure~\ref{fig:bifpinnv} shows a comparison of vertical slices through the center of the cube and in the middle of the time series. The physical parameters are reasonably well reproduced around $z=0$ where most of the continuum radiation originates. Above that and in the upper photosphere, our radiative loss approximation is not valid, and we do not take into account radiative heating. Below the height where most radiation is emitted, there are more discrepancies, most likely due to the fact that we do not have sufficient information to constrain the flows below the visible surface; the upflows and downflows tend to be mostly vertical in the PINN, while the Bifrost model shows much more complicated structures. \deleted{One may wonder how the real Sun looks like and whether the structures in the Bifrost simulations are partly due to the lower boundary.} Furthermore, the sharp boundaries in density and temperature are not well reproduced due to the inability of the current neural network to represent sharp boundaries. Note that a denser sampling of points where the constraints are enforced cannot overcome the spectral bias; potential solutions are discussed below.

\begin{figure}
  \includegraphics[width=0.48\textwidth]{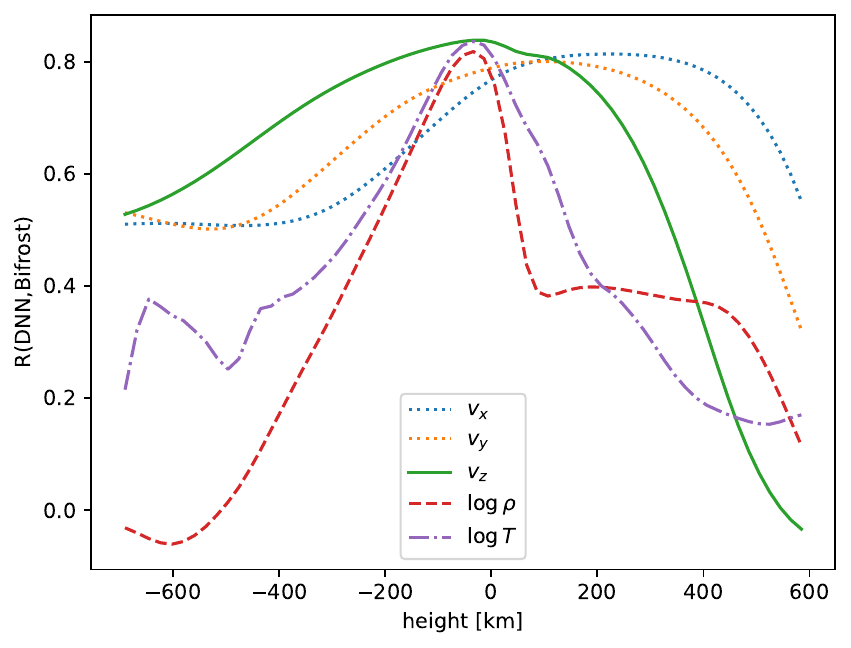}
  \caption{Pearson correlation coefficients as a function of height between the Bifrost model and the PINN for the three velocity components and the logarithms of the density and the temperature. The height scale has been adjusted so that $z=0$ corresponds to the average height were the optical density at 500~nm reaches unity. The correlation coefficient is high just below that level where most of the continuum radiation is formed.}
  \label{fig:corrpinn}
\end{figure}

Figure~\ref{fig:corrpinn} shows the Pearson correlation coefficients between the Bifrost simulation and the PINN output for the five physical parameters as a function of height. The correlation clearly peaks where most of the continuum radiation emerges. The velocities show a significant correlation below and above that level; this is not surprising as the velocities are reasonably continuous across the layer where the majority of the radiative loss occurs (see Fig.~\ref{fig:bifpinnv}). The correlation coefficients of the density and temperature drop much more rapidly above the $\tau=1$ level, which we attribute to the PINN not being able to reproduce the sharp transition just below the $\tau=1$ surface as well as our extremely simplified radiative loss calculation, which excludes radiative heating and cannot reproduce the radiative losses due to strong spectral lines in higher layers.

\begin{figure}
  \includegraphics[width=0.48\textwidth]{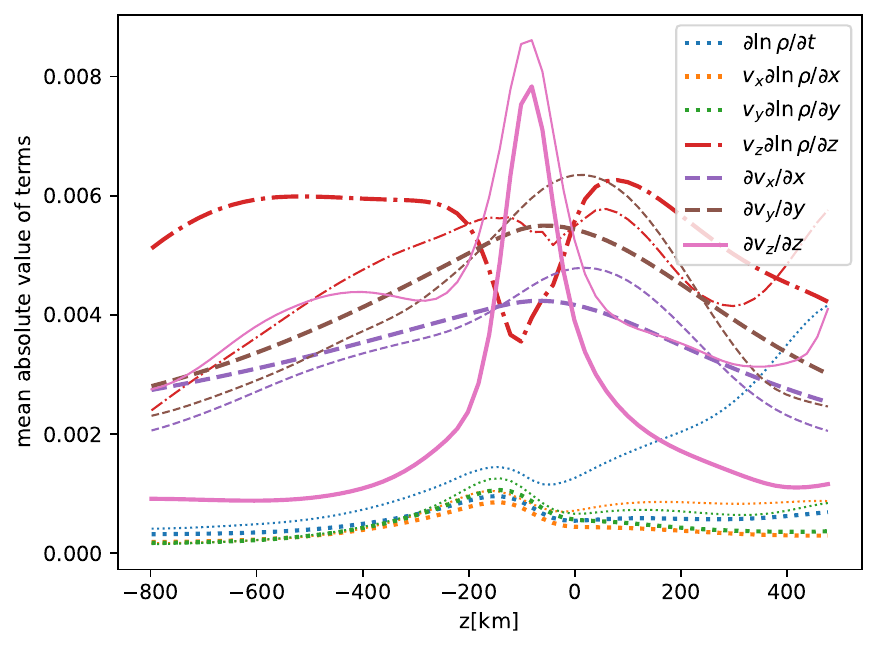}
  \caption{Horizontally averaged absolute magnitudes of all terms of the continuity equation as a function of height in the PINN (thick lines) and the corresponding Bifrost simulation (thin lines).}
  \label{fig:pinn_cont}
\end{figure}

\begin{figure}
  \includegraphics[width=0.48\textwidth]{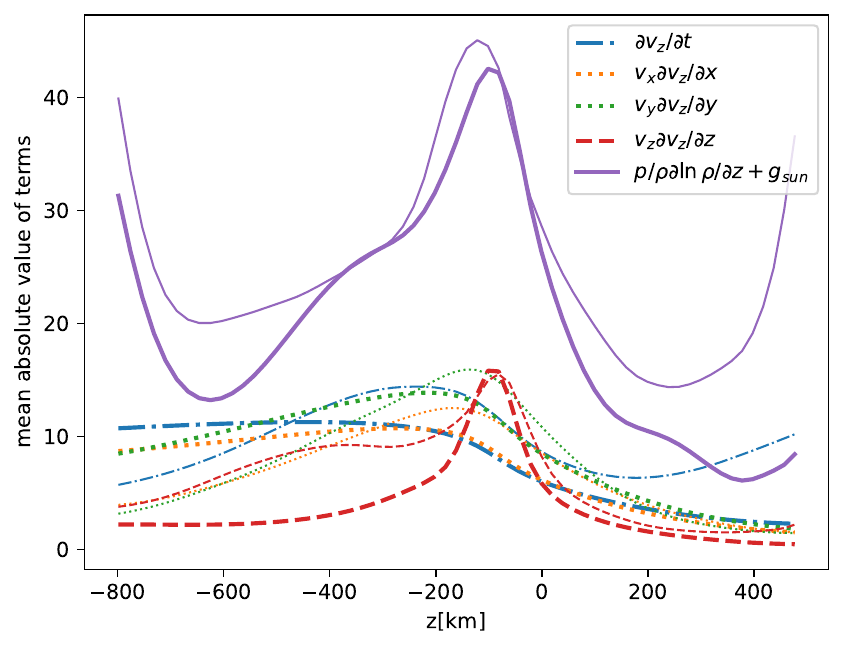}
  \caption{Horizontally averaged absolute magnitudes of all terms of the vertical momentum equation as a function of height in the PINN (thick lines) and the corresponding Bifrost simulation.}
  \label{fig:pinn_momv}
\end{figure}

\begin{figure}
  \includegraphics[width=0.48\textwidth]{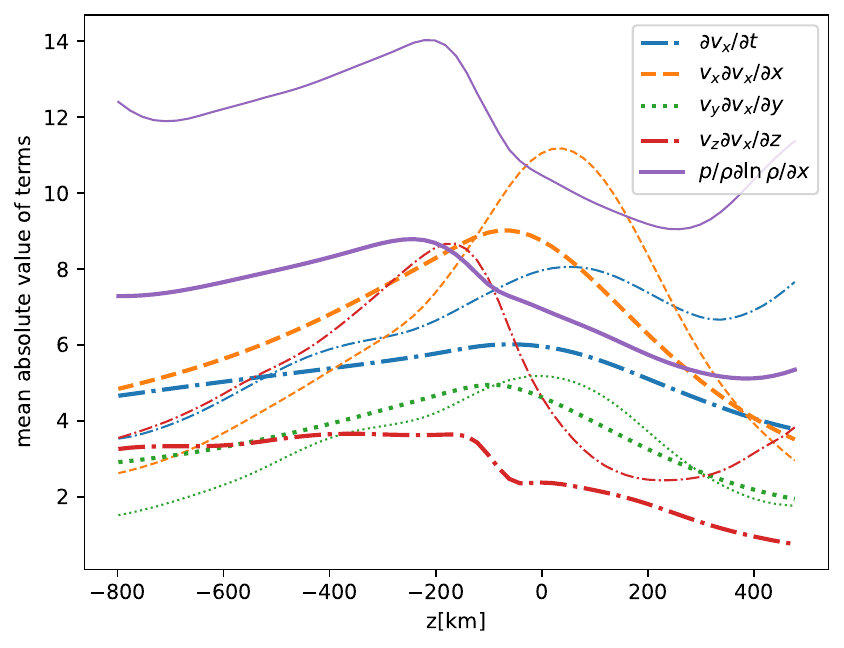}\hfill
  \includegraphics[width=0.48\textwidth]{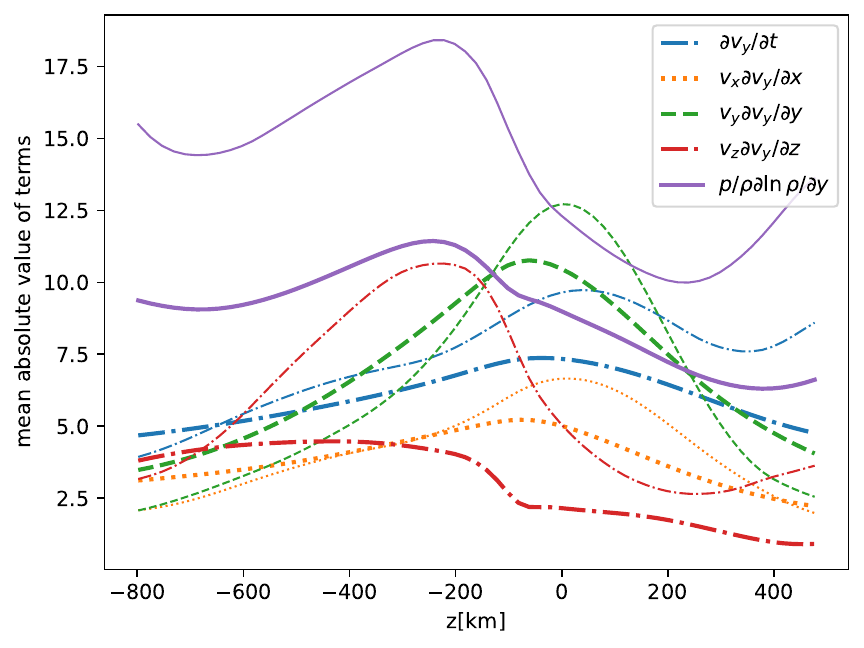}
  \caption{Horizontally averaged absolute magnitudes of all terms of the horizontal momentum equations as a function of height in the PINN (thick lines) and the corresponding Bifrost simulation.}
  \label{fig:pinn_momh}
\end{figure}

\begin{figure}
  \includegraphics[width=0.48\textwidth]{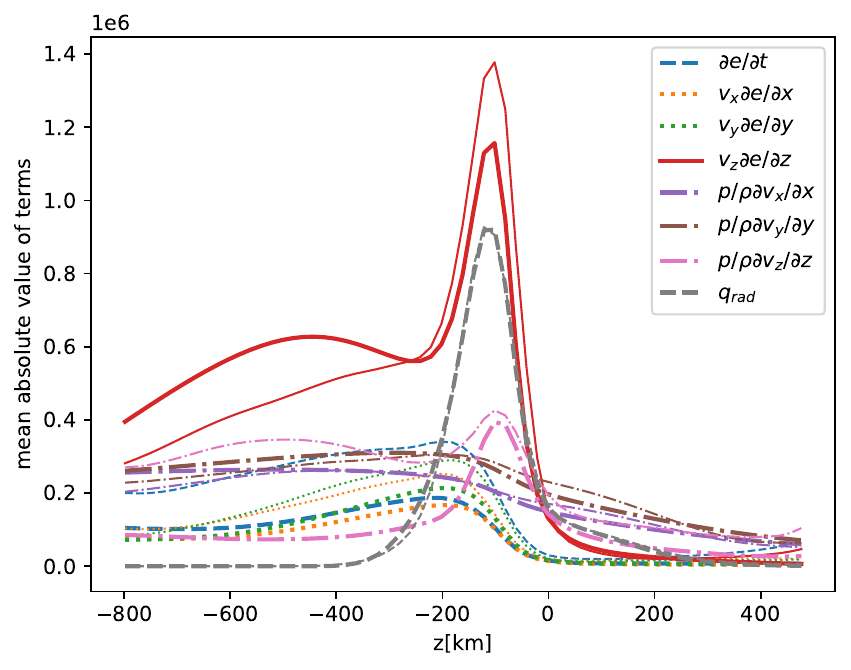}
  \caption{Horizontally averaged absolute magnitudes of all terms of the energy equation as a function of height in the PINN (thick lines) and the corresponding Bifrost simulation.}
  \label{fig:pinn_ener}
\end{figure}

\added{
  We may also look at the different terms in the PDEs to see how well the PINN recovers them. Figures~\ref{fig:pinn_cont} to \ref{fig:pinn_ener} compare these terms between the PINN in thick lines and the corresponding Bifrost simulation in thin lines. Most terms in the continuity equation (Fig.~\ref{fig:pinn_cont}) are well represented by the PINN in the layers where the continuum radiation originates except for the terms that involve the derivative with respect to the vertical $z$-axis. This is not too surprising as the PINN struggles with reproducing the sharp transition there. The vertical momentum equation terms (Fig.~\ref{fig:pinn_momv}) are all surprisingly well represented, most likely because of the tight coupling between the vertical velocity and the intensity around optical depth unity. Interestingly, the horizontal momentum equation terms (Fig.~\ref{fig:pinn_momh}) show significant discrepancies that are highly consistent between the two horizontal directions. This may be related to the difficulty of accurately retrieving horizontal velocities and having to actively bias against unreasonably large horizontal velocities in the horizontal momentum constraints (see Sect.~\ref{subsec:momentum}). Finally, the terms of the energy equation (Fig.~\ref{fig:pinn_ener}) are all reasonably well represented; the term involving the derivative of the inner energy per mass with respect to the vertical direction shows some difference, again likely due to the inability of the PINN to represent very sharp transitions.
}

\begin{figure}
  \includegraphics[width=\textwidth]{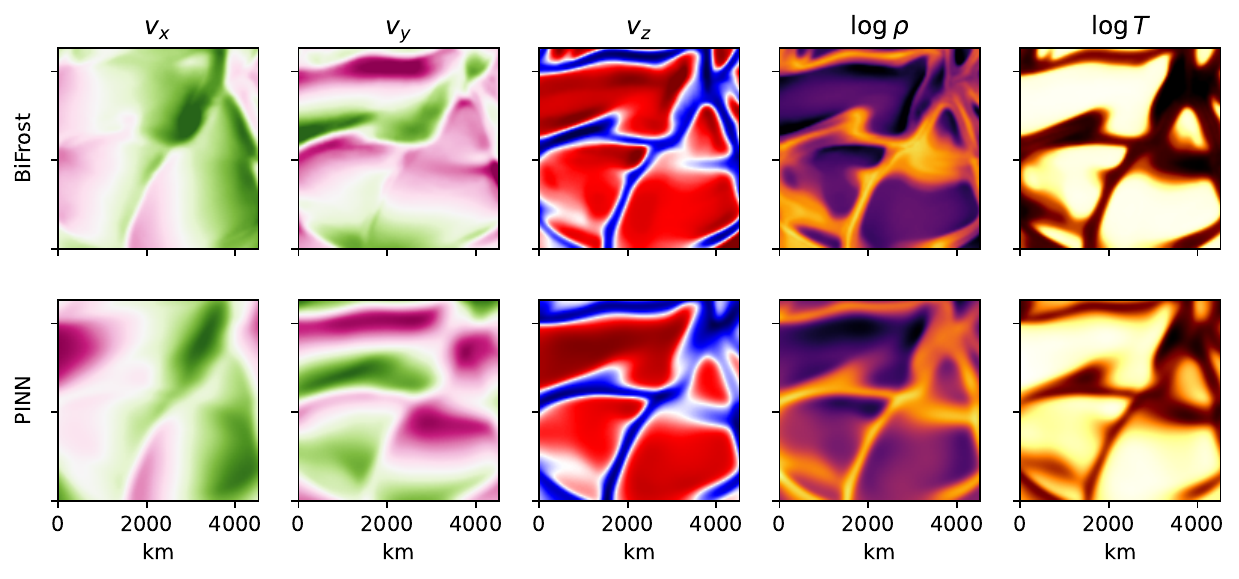}
  \caption{Same as Fig.\ref{fig:bifpinn} but for a time step 30 seconds later where a Bifrost snapshot was available but no constraints were applied, i.e.\ no image was provided for this time and no PDE constraints were enforced. This shows that the PINN does not produce unreasonable values between times when observational and physics constraints have been applied.}
  \label{fig:interpol}
\end{figure}

There are other tests that one might consider including interpolations and extrapolations. For an interpolation test, one can compare the PINN results with Bifrost simulation time steps that have not been used in the generation of the input images; there are four Bifrost time steps available between every one of the snapshots that we used to generate the images. \replaced{So far,}{Figure~\ref{fig:interpol} shows the same as Fig.~\ref{fig:bifpinn} but for a Bifrost snapshot 30 seconds after the one in Fig.~\ref{fig:bifpinn}; no observational or physics constraints were applied at that time step when the PINN was trained. This shows that the PINN does not produce unreasonable results between points in time where constraints were applied. However,} the PINN-based images are not doing significantly better than a linear interpolation of the images that were used to train the PINN. This is good news as one might have wondered whether the PINN becomes unrealistic between the times when images are available. The lack of a significant improvement over a linear interpolation is likely due to the fact that our time-step of 50~s is too long, and the physics constraints are not applied frequently enough.

Extrapolation tests failed when we trained the PINN on the first half of the images from the Bifrost simulation and then kept enforcing the physics constraints while dropping the image constraint; a very bright feature formed at the edge along with larger, lower-contrast granules in the rest of the area. This is likely due to the absence of any kind of boundary conditions on the sides or constraints regarding the horizontal mass and energy flows. These interpolation and extrapolation tests can also be easily implemented when using observational data by only using a part of the observational data.

\section{Discussion} \label{sec:discussion}

\subsection{Advantages and disadvantages of PINN approach}
The PINN approach to creating data-driven models of the solar atmosphere has many distinct advantages:
\begin{itemize}
  \item Inherent simplicity: the physics is implemented directly as partial differential equations, which can typically be done in a single line of code per equation, and the calculation of synthetic observations can largely be copied from existing codes. As such, it is easy to add additional physics and/or observations.
  \item Optimal compression: compared to traditional, grid-based schemes to solve PDEs, the PINN approach uses an almost maximally compressed version of the information\added{, with the major caveat that the compression is lossy and that the PDEs are only approximately fulfilled}. In this proof of concept using Bifrost simulations, the amount of information required to represent the time-dependent, three-dimensional atmosphere is reduced by two to three orders of magnitude. This enables the use of GPUs that typically have less memory available than CPUs, and small-scale problems like the one explored here can even be computed on a powerful laptop. The compression ratio is likely to strongly depend on the numerical simulations that are being used.
  \item Balance of accuracy and precision: classical inversions and numerical solutions of PDEs tend to be exceedingly precise but their accuracy may be limited by the errors in the observational data in the case of inversions and/or the approximation of the physics. The extreme precision in numerical solvers is not balanced with the limited accuracy of the data and/or the physics, and there is no simple approach to reduce the precision and gain in terms of computing time. PINNs on the other hand, provide a simple way to balance precision and accuracy as the precision increases as the optimization progresses; the longer the optimization runs, the better the PINN obeys the PDEs. There is no need to do everything with the utmost precision when the observational data have limitations.
  \item No grid: The PINN model is continuous in space and time and can be evaluated at arbitrary locations without the need for interpolation. This opens the possibility to zoom into details such as as bright points, focus on regions of interest such as the sharpest parts of images where adaptive optics provides the best wavefront correction, represent a particular layer of the atmosphere with great detail while approximating lower levels, or focus on an interesting point of time when a particular event happens such as a flare. And derivatives can be calculated anywhere with little computational cost thanks to the analytical nature of neural networks that enables automatic differentiation. Finally, different PDEs can be enforced on their natural spatial and temporal scales; there is is no need to calculate everything at the smallest and/or fastest scale.
  \item Simple integration of observational data: PINNs make it easy to solve the PDEs with observational data as boundary conditions because the solution of the PDEs and the matching of the observational data are both done by optimizing the free parameters of a neural network with both the PDEs and the observational data as boundary conditions. There is no requirement for the observational data to be synchronized, equally spaced in space or time, or on the same spatial scales. This is particularly interesting when combining data from different observatories and/or instruments. It even makes it easy to integrate data from a scanning spectrograph since the individual spectra, taken at different times, can easily be used as observational constraints. In the future, even the Point Spread Function (PSF) may be included to simulate the limited spatial resolution of observations; this is particularly easy to integrate as many neural networks include convolution operations. It may even be possible to infer the PSF from the data when the actual PSF of the observations is not well determined.
  \item Progressive complexity: PINNs make it very easy to progressively increase the complexity of the physics that is considered in the model. For instance, one can easily switch from the anelastic approximation to the full continuity equation by changing a single line in the code. When increasing the complexity in the physics, one does not have to start from scratch but can use the PINN from the simpler approximation as the starting point for training a PINN with more elaborate physics. In the future, it should be possible to first train a hydrodynamic model before increasing the complexity and adding magnetic fields.
\end{itemize}


\subsection{Limitations of current proof of concept} \label{subsec:limitations}
The current effort shows that the concept is feasible: PINNs can create physically consistent, time-dependent radiative hydrodynamics models that explain observations even if the only observational constraint is a sequence of continuum images. There are many parts where the current effort is too simplistic such as the radiative loss calculation and the equation of state. In the following we will discuss some of the limitations that concern more fundamental aspects of the approach and that one can expect to face in the future even if the physics complexity is increased to the level where it matches current simulation codes such a Bifrost. Potential solutions will be discussed in Sect.~\ref{sec:outlook}.

\begin{itemize}
\item Temporal physics sampling: Our spatial sampling of the physical constraints is equivalent to the original Bifrost data, which should be sufficient to resolve the relevant spatial scales of the Bifrost simulations. However, our time step is about 100 times longer than what is typically used by Bifrost in the photosphere. As such, it may be that our PINN provides physically self-consistent snapshots, but the snapshots may correspond to different trajectories of PDE solutions. For simple PDEs such as a harmonic oscillator, the different trajectories are often parametrized with conserved quantities such as the total energy. While we have tried to incorporate some of this approach by keeping the total energy flux fixed at all heights and requiring no net mass flux through any horizontal layer, it is not clear how this impacts the choice of time step. It will be necessary to develop the equivalent of the Courant et al.\ (\citeyear{Courant1928}) criterion to establish the necessary temporal density in enforcing the physics constraints. 
\item Uniqueness: There is no guarantee that the PINN model solution is unique or that the optimizer has found the global minimum; it is actually highly unlikely. One might start from different starting points to obtain an idea of the range of solutions that are compatible with the observations and the physics, but it is not obvious that this will be sufficient to understand what aspects of the returned model can be considered to be unique. 
\item Unknown errors: Classical optimization approaches such as Levenberg-Marquardt \citep{Levenberg1944, Marquardt1963} or Markov Chain Monte Carlo \citep[e.g.][]{GamermanLopes2006} provide information on the error and correlations of free parameters that are fitted, but they tend to be limited to a small number of parameters. The optimizers used for neural network training can deal with millions to billions of free parameters, but they typically do not provide any information on the posterior distribution of the model parameters. 
\item Unknown neural-network viscosity: Deep Neural Networks are better at reproducing low-frequency variations than high-frequency variations, a phenomenon known as spectral bias \citep{Rahaman2018}. As such, the deep neural network will dampen small-scale variations in the atmospheric parameters; while one can evaluate the damping by looking at Fourier transforms of parameters, this neural-network viscosity cannot be adjusted in an easy way.
\item Scalability: The current proof of concept has been developed on a relatively small area of the simulation covering about two orders of magnitude in spatial frequencies. One might consider working with a much larger surface area and increase the number of free parameters of the neural network by two orders of magnitude or so; some quick tests did not succeed because the memory requirements could not be accommodated, and tests on slightly larger scales than shown here that still fit into memory suffered too much from the spectral bias.
\item No boundary conditions: The current implementation does not enforce any type of boundary conditions. While constraining the mean vertical density and temperature stratifications, enforcing a fixed total vertical energy flow at all heights and requiring zero net mass motion at any height limits the range of potential solutions that are compatible with the observations, it does not enforce specific boundary conditions at the top and bottom, and most certainly not on the sides. The later may be the reason that experiments to realistically extrapolate convective motions in time was not successful. 
\end{itemize}

\section{Outlook} \label{sec:outlook}

The proof of concept outlined here neglects many important physical processes and makes overly simplified approximations and assumptions. The equation of state can easily be improved; tabular values from comprehensive models can be used to train a neural network in analogy to what we did for the opacities. Improving the radiative transfer calculations is not trivial as that requires non-local values, which are more difficult to calculate within the PINN framework. One might consider using an amended or separate neural network to carry out the integration along different orientations. In the simplest case of the optical depth in the vertical direction, one might choose to have the optical depth as an additional output quantity of the PINN; two additional constraints could then be used to train the PINN to provide the correct optical depth: 1) $\partial\tau/\partial z = \kappa\rho$ with $\kappa$ being the opacity per unit mass and 2) the optical depth having a fixed value at a certain height such as being zero at the top of the box. Attempts to use this approach were unstable during the optimization. 

To be generally useful, magnetic fields will need to be added, and (polarized) spectral line information must be used as observational constraints. Furthermore, radiative losses in higher layers need to be added, and waves and oscillations need to be considered.

To estimate the uncertainty in the retrieved physical parameters, one might consider using a Bayesian approach were one fits the distribution of physical parameters at each location instead of the parameter itself. Such Bayesian neural networks \citep[e.g.][]{Jospinetal2020} could provide a way to estimate the error, but this may require unrealistically high computing power.

The artificial viscosity introduced by the deep neural network can probably be understood in terms of the Neural Tangent Kernel \citep{Jacotetal2018}. To control it, large language models that are all the rage in AI development right now may show a way forward with positional encoding \citep{Vaswanietal2017}. Translated to the PINN approach, instead of feeding the $(x,y,z,t)$ position vector into the neural network, one feeds sines and cosines of suitably scaled positions in the four axes into the neural network. An initial attempt at implementing this approach in our PINN failed miserably; while small-scale structures were recovered with much improved accuracy, the solutions between points where the physics and observational constraints was enforced, were completely unrealistic. Maybe there are suitable ways to add positional encoding such that the artificial viscosity can be controlled. 

When it comes to applying the PINN approach to larger fields of view, it may not be the best approach to simply work with a larger field directly but split the area into smaller fields, with individual PINNs working only on the smaller fields. The Finite Basis Physics-Informed Neural Networks \citep{Mosleyetal2021} may provide the basis for successfully dealing with the scale issue: a smooth, differentiable window function locally confines each PINN to its corresponding subdomain. In addition, one may also want to optimize the points where the PDE constraints are enforced, and those locations may depend on the particular PDE or physics it implements. For instance, one might concentrate points in space and time where changes occur on short spatial or temporal scales.

The research presented here shows that PINNs may be a viable approach for data-driven modeling of the solar atmosphere. It is only the very first step in an exciting direction that has the potential to revolutionize the way we interpret solar observations, understand the underlying physics and approximate solar processes on small scales such that they can be efficiently included in simulations at much larger scales.


\begin{acknowledgments}
Matthias Rempel provided most helpful comments and educated me on some of the intricate aspects of modeling the solar photosphere. I thank the team at Lowell Observatory for providing an environment where I could work myself into a new topic and Agaath van Dorp and David Giancoli for their hospitality while I finished the manuscript. \added{The National Solar Observatory is operated by the Association of Universities for Research in Astronomy, Inc.\ (AURA), under cooperative agreement AST-1400450 with the US National Science Foundation.}
\end{acknowledgments}

\software{
TensorFlow (Abadi et al. 2015),
Keras (Chollet et al. 2015),
numpy (Harris et al. 2020),
astropy (The Astropy Collaboration 2013, 2018, 2022),
matplotlib (Hunter 2007)
}

\bibliography{rhpinn_arxive}{}
\bibliographystyle{aasjournal}

\end{document}